\documentclass{aa}
\usepackage{graphicx,natbib,epsfig}

\def\mearth{M_\oplus}
\begin{document}
\title{Saturn's internal structure and carbon enrichment}

\author{O. Mousis \inst{1}, Y. Alibert \inst{2} and W. Benz\inst{2}}

\institute{Observatoire de Besan\c{c}on, CNRS-UMR 6091, BP 1615, 25010 Besan\c{c}on Cedex, France \and Physikalisches Institut, University of Bern, Sidlerstrasse 5, CH-3012 Bern, Switzerland\\
             }

\offprints{O. Mousis, \\   e-mail: Olivier.Mousis@obs-besancon.fr}

\date{Received / Accepted}

\titlerunning{Volatiles enrichments in Saturn}

\abstract{We use the clathrate hydrate trapping theory to calculate the  enrichments in O, N, S, Xe, Ar and Kr compared to solar  in Saturn's atmosphere. For this, we calibrate our calculations using two different carbon  abundance determinations that cover the domain of measurements published in the last decades: one derived from the NASA $Kuiper$ Airborne Observatory measurements and the other obtained from the {\it Cassini} spacecraft observations. We show that these two different carbon abundances imply quite a different minimum heavy element content for Saturn. Using the $Kuiper$ Airborne Observatory measurement for calibration, the amount of ices accreted by Saturn is found to be consistent with current interior models of this planet. On the other hand, using the {\it Cassini} measurement for calibration leads to an ice content in the planet's envelope which is higher than the one derived from the interior models. In this latter case, reconciling the interior models with the amount of C measured by the {\it Cassini} spacecraft requires that significant differential sedimentation of water and volatile species have taken place in Saturn's interior during its lifetime.


}

\maketitle

\section{Introduction}

The standard giant planet formation scenario, namely the so-called core accretion model (Pollack et al. 1996),  has been recently extended by Alibert et al. (2004; 2005a,b) to include migration of the growing planet  and proto-planetary disk evolution. These calculations show that the extended core accretion model can lead to the formation of two giant planets closely resembling Jupiter and Saturn in less than 3 
Myr (Alibert et al. 2005b, hereafter A05b). Based on these formation models, Alibert et al. (2005c, hereafter A05c) also demonstrated that the volatiles enrichments measured in Jupiter's atmosphere can be explained  in a way consistent with the constraints set from internal structure modeling (Saumon \& Guillot 2004, hereafter SG04). In this work, we follow the strategy adopted by A05c and utilize the thermodynamical conditions of the protoplanetary disk that led to the formation of Saturn (A05b) to calculate the volatile enrichments in its atmosphere. We then compare the calculated amount of required ices in Saturn with the content in heavy elements predicted by the internal structure models of SG04.\\

The carbon abundance in Saturn's atmosphere has been the subject of a large number of studies. In our calculations, we adopt two extreme values of C abundance that cover the domain of measurements published in the last 25 years (see Table \ref{CH4}). From the use of a haze model to derive volume mixing ratios in the near-IR bands, C has been shown to be enriched by a factor of $3.4 \pm 0.9$ (Kerola et al. 1997; $Kuiper$ Airborne Observatory (KAO)) in Saturn's atmosphere compared to its solar value\footnote{Note that we use the central values of species abundances derived from Table \ref{table_L03} of Lodders (2003) in our calculations, contrary to A05b and A05c who adopted a set of values within the error bars to obtain the best fit possible of the volatile enrichments in Jupiter.} (Lodders 2003;  see Table \ref{table_L03}). On the other hand, recent {\it Cassini} CIRS far-IR measurements lead Flasar et al. (2005)  to infer a C enrichment of $8.8 \pm 1.7$ compared to solar from the fit of CH$_4$ rotational lines.

Moreover, from an indirect determination, nitrogen has been shown to be enriched by a factor of $3.1 \pm 0.6$ (Briggs \& Sackett 1989; ground-based radio wavelength observations) in Saturn's atmosphere compared to solar. Briggs \& Sacketts (1989) also estimated that S is enriched by a factor of $\sim$ 12 compared to solar in Saturn. However, this latter estimation seems to be uncertain due to the difficulty of distinguishing in the microwave spectrum between the contribution to the opacity of H$_2$S from that of other compounds (Hersant et al. 2004). For this reason, we assume in the present work that the abundance of S is still unknown.

\begin{table*}[h]
\caption[]{List of measurements of CH$_4$ abundances in Saturn's atmosphere published since 1981.}
\begin{center}
  \begin{tabular}{ccccc}
            \hline
            \hline
          \noalign{\smallskip}
    CH$_4$/H$_2$ ($\times$ $10^{-3}$) & Solar enrichment &  Reference & Measurement  \\
   \noalign{\smallskip}
  \hline
    \noalign{\smallskip}
    2 $\pm$ 0.5 & 3.4 $\pm$ 0.9 & Kerola et al. (1997) & KAO  \\ 
    $2^{+1.0}_{-0.8}$ & $3.4^{+1.7}_{-1.4}$ & Encrenaz \& Combes (1982) & ground-based \\ 
    $2.2^{+0.8}_{-0.2}$ & $3.8^{+1.4}_{-0.3}$ & Killen (1988) & ground-based  \\ 
    2.5 & 4.3 & Tomasko \& Doose (1984) & $Pioneer$ IPP  \\ 
    3 $\pm$ 0.6 & 5.2 $\pm$ 1.0 & Karkoschka \& Tomasko (1992) & ground-based  \\ 
    4 $\pm$ 2 & 6.9 $\pm$ 3.4 & Buriez \& de Bergh (1981) & ground-based \\ 
    4.2 $\pm$ 0.4 & 7.2 $\pm$ 0.7 & Trafton (1985) & ground-based  \\ 
    4.4 $\pm$ 1.2 & 7.6 $\pm$ 2.1 & Lellouch et al. (2001) & ISO-SWS  \\ 
   $4.5^{+2.4}_{-1.9}$ & $7.7^{+4.1}_{-3.3}$ & Courtin et al. (1984) & $Voyager$ IRIS  \\ 
    5.1 $\pm$ 1.0 & 8.8 $\pm$ 1.7 & Flasar et al. (2005) & $Cassini$ CIRS  \\ 
  \hline
     \end{tabular}
     \end{center}
\label{CH4}
\end{table*}

\section{Formation of Saturn}

The model of Saturn's formation considered here is the one presented by A05b. In this model, Saturn forms from an embryo originally located at $\sim$ 12 AU. The proto-planet migrates inwards by following proto-Jupiter's trail and stops at the current position of Saturn when the disk disappears. Planetesimals are accreted during the whole formation process, along proto-Saturn's migration pathway (between 12 and 9.8 AU). The resulting planet exhibits an internal structure comparable to that of the actual Saturn described by SG04\footnote{Detailed internal structure models of Saturn developed by SG04 indicate a maximum mass of heavy elements present in the envelope between nearly 0 and 10 $\mearth$, whereas the mass of the core varies between 8 and 25 $\mearth$. 
However, note that the minimum mass of the solid core might be decreased by up to $\sim$ 7 $\mearth$ depending upon the extent of sedimented helium.}, with a final core mass of 
$\sim$~6~$\mearth$ and a total mass of accreted planetesimals of $\sim$~13.2~
$\mearth$.  Finally, the thermodynamical conditions in the disk at the actual position of Saturn are used do calculate the composition of planetesimals that were accreted by the planet.

\begin{figure}
\centerline{\includegraphics[width=6cm,angle=-90]{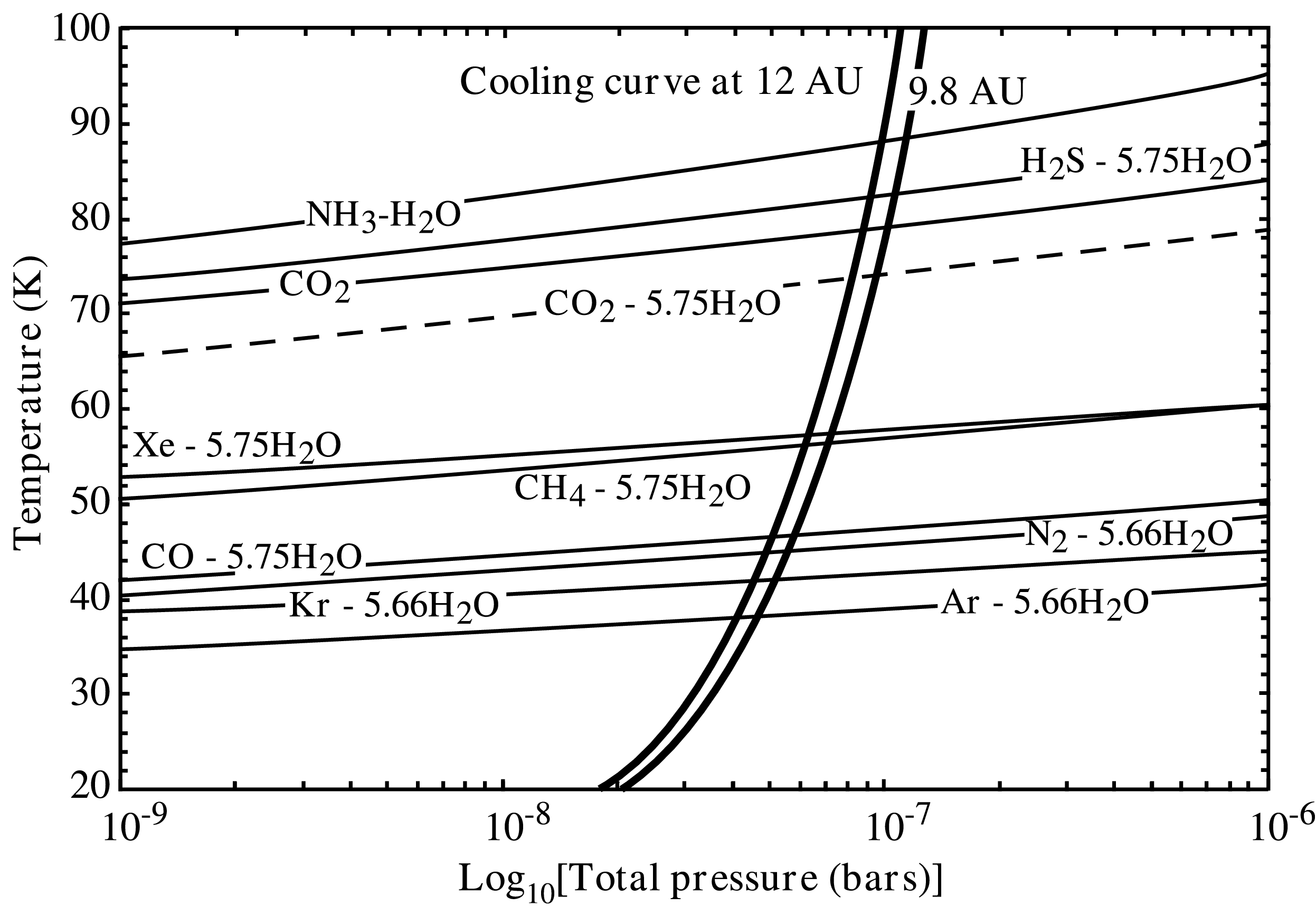}}
\caption{Stability curves of the condensates considered in the present work, and evolutionary tracks of the nebula at 12 and 9.8 AU. Abundances of various elements are solar, with CO$_2$:CO:CH$_4$ = 30:10:1 and N$_2$:NH$_3$ = 1 in vapor phase. The condensation curve of CO$_2$ pure condensate (solid line) is shown together with that of the corresponding clathrate hydrate (dashed line). Species remain in the vapor phase as long as they stay in the domains located above the curves of stability.}
\label{cooling}
\end{figure}

\section{Composition of planetesimals in Saturn's feeding zone}

In this work, we consider that abundances of all elements are solar and that O, C, and N exist only under the form of H$_2$O, CO$_2$, CO, CH$_4$, N$_2$, and NH$_3$ in the solar nebula gas-phase. Concerning the initial ratios of CO$_2$:CO:CH$_4$ and N$_2$:NH$_3$ in the solar nebula, for the sake of consistency, we adopt the same ranges of values as the ones derived by A05c to fit the observed enrichments in volatiles in Jupiter. Hence, we assume CO$_2$:CO:CH$_4$~=~40:10:1 to 10:10:1 for N$_2$:NH$_3$~=~1 and CO$_2$:CO:CH$_4$ = 10:10:1 for N$_2$:NH$_3$~=~10 in the vapor phase of the solar nebula. The considered values of CO$_2$:CO:CH$_4$ are consistent with ISM measurements (Allamandola et al. 1999; Gibb et al. 2004). The adopted values of N$_2$:NH$_3$ are compatible with thermochemical calculations in the solar nebula that take into account catalytic effects of Fe grains on the kinetics of N$_2$ to NH$_3$ conversion (Fegley 2000). In addition, S is assumed to exist under the form of H$_2$S (H$_2$S/H$_2$ = 0.7 $\times$ (S/H$_2$)$_\odot$, following A05c) and other sulfur compounds in the solar nebula (Pasek et al. 2005). 

Finally, we also assume that volatiles have been trapped during the cooling of the nebula in planetesimals either in form of pure condensates, as it is the case for CO$_2$, or in form of hydrates or clathrate hydrates, as it is the case for NH$_3$, N$_2$, CH$_4$, CO, Xe, Ar and Kr (see A05c for details). Note that the amount of available water ice is assumed to be sufficient to trap all volatiles except CO$_2$. Once condensed, ices are assumed to decouple from the gas and to be incorporated into growing planetesimals which may subsequently be accreted by the forming Saturn.

Figure \ref{cooling} represents the cooling curves of the nebula at 9.8 and 12 AU derived from the solar nebula model used by A05b, as well as the condensation curves for the various ices considered in this work. The stability curves of clathrate hydrates are derived from Lunine \& Stevenson (1985) whereas we used a fit to the experimental data for the pure CO$_2$ condensate (Lide 1999). Figure \ref{cooling} provides the condensation sequence of the different volatiles initially existing in vapor phase inside Saturn's feeding zone. The intersection between the cooling curve and the stability curve of the different condensates also gives the thermodynamical conditions at which the different ices are formed. 

Using Eqs. 1 and 2 given in A05c, we can calculate 1) the minimum abundance of water required to trap all volatiles (except CO$_2$) in the solar nebula gas-phase to form clathrate hydrates or hydrates and 2) the composition of ices incorporated in planetesimals that were accreted by proto-Saturn. Since, as shown in Table \ref{planetesimaux}, the composition of planetesimals does not vary significantly along the migration path of Saturn, we assume that all the ices contained in accreted planetesimals have an identical composition.

\begin{table}[h]
\caption[]{Gas phase abundances (molar mixing ratio with respect to H$_2$) of major species in the
solar nebula (from Lodders 2003) for CO$_2$:CO:CH$_4$~=~30:10:1 and N$_2$:NH$_3$~=~1.}
\begin{center}
\begin{tabular}[]{lclc}
\hline
\hline
\noalign{\smallskip}
Species $i$ & $x_i$ & Species $i$ & $x_i$\\
\noalign{\smallskip}
\hline
\noalign{\smallskip}
O &   $1.16 \times 10^{-3}$& N$_2$  &   $5.33 \times 10^{-5}$  \\
C &  $5.82 \times 10^{-4}$  & NH$_3$  &  $5.33 \times 10^{-5}$ \\
N &  $1.60 \times 10^{-4}$ & S & $3.66 \times 10^{-5}$\\
H$_2$O  & $1.66 \times 10^{-4}$ & Ar & $8.43 \times 10^{-6}$ \\
CO$_2$  & $4.26 \times 10^{-4}$ & Kr & $4.54 \times 10^{-9}$ \\
CO  & $1.42 \times 10^{-4}$  & Xe  & $4.44 \times 10^{-10}$ \\
CH$_4$  & $1.42 \times 10^{-5}$ \\
\hline
\end{tabular}
\end{center}
\label{table_L03}
\end{table}

\begin{table}[h]
\caption[]{Calculations of the ratios of trapped masses of volatiles to the mass of 
H$_2$O ice in planetesimals formed at 9.8 and 12 AU in the solar nebula. Gas-phase abundance of H$_2$O is equal to $8.11 \times 10^{-4}$ ($\sim$ 4.9 times the value quoted in Table \ref{table_L03}; see text), and gas-phase abundances of elements are assumed to be solar (Lodders 2003) with CO$_2$:CO:CH$_4$~=~30:10:1 and with N$_2$:NH$_3$~=~1 in vapor phase in the solar nebula. 
The abundance of H$_2$S is subsolar (see text).}
         \begin{center}
         \begin{tabular}[]{lcc}
            \hline
            \hline
            \noalign{\smallskip}
               &  9.8 AU & 12 AU  \\
             \hline
             \noalign{\smallskip}
             CO$_2$:H$_2$O  &  $8.91 \times 10^{-1}$ &  $9.13 \times 10^{-1}$  \\
             CO:H$_2$O  &  $1.46 \times 10^{-1}$ &  $1.45 \times 10^{-1}$  \\
             CH$_4$:H$_2$O     &  $9.14 \times 10^{-3}$  &  $9.23 \times 10^{-3}$  \\
             N$_2$:H$_2$O      &  $5.26 \times 10^{-2}$  &  $5.34 \times 10^{-2}$   \\
             NH$_3$:H$_2$O    &  $4.67 \times 10^{-2}$   &  $4.66 \times 10^{-2}$  \\
             H$_2$S:H$_2$O     &  $4.28 \times 10^{-2}$  &  $4.32 \times 10^{-2}$  \\
             Ar:H$_2$O    &  $9.74 \times 10^{-3}$  &  $9.94 \times 10^{-3}$  \\
             Kr:H$_2$O    &  $1.34 \times 10^{-5}$  &  $1.34 \times 10^{-5}$  \\
             Xe:H$_2$O    &  $2.44 \times 10^{-6}$  &  $2.45 \times 10^{-6}$  \\
            \hline
         \end{tabular}
         \end{center}
         \label{planetesimaux}
         \end{table}

\section{Enrichments in volatiles in Saturn}

From the determination of the composition of the ices incorporated in the planetesimals which depend on the initial gas-phase composition in the solar nebula and the adjustment of the amount of heavy elements accreted by proto-Saturn, it is possible to calculate the expected enrichments in volatiles in the giant planet's atmosphere. In Table \ref{enrichissements} we present the results of these  computations and compare them to the observed enrichments for CO$_2$:CO:CH$_4$~=~30:10:1 and N$_2$:NH$_3$~=~1 (our nominal vapor phase conditions) in the solar nebula gas-phase. It can be seen that our model reproduces the C and N ground-based enrichment measurements. The minimum fit of the airborne C and ground-based N measurements corresponds to an O enrichment of 5.4 times the solar value. From this, we predict that S, Ar, Kr and Xe are overabundant compared to solar by a factor of 2.1, 2.0, 2.2 and 2.6 respectively. In addition, these enrichments result from the accretion of 5.6 $\mearth$ of ices including 2.5 $\mearth$ of water in the envelope of Saturn. The resulting total content in heavy elements in Saturn is therefore compatible with the formation model of A05b and consequently with the internal structure model of SG04, provided the ratio of ices to rocks (I/R) is greater or equal to $\sim$ 0.7. Table \ref{ices} summarizes the ranges of required masses of ices in the envelope of Saturn, and the resulting O enrichments, to match both C and N ground-based measurements if the initial gas phase conditions are varied. It can be seen that O is enriched at least between 5.1 and 7.6 times the solar abundance in Saturn's atmosphere while the minimum mass of incorporated ices varies between 5.4 and 7.6 $\mearth$ in the giant planet. Under these conditions, the minimum I/R ratio required in Saturn varies between $\sim$ 0.7 and 1.4 in order to be compatible both with the formation model of A05b and the internal structure model of SG04.

On the other hand, fitting the recent C measurement derived from {\it Cassini} data does not allow to retrieve a value of N compatible with the measured one. In the nominal case, fitting the minimum value of the {\it Cassini} measurement implies overabundances compared to solar of N, S, Ar, Kr and Xe by factors of 6.5, 5.5, 5.1, 5.6 and 6.7 respectively.  These calculations translate to an O enrichment of at least 13.9 times the solar value in the planet's  envelope and require the accretion of at least 14.4 $\mearth$ of ices including 6.6 $\mearth$ of water. Following Table \ref{ices}, if the initial gas-phase conditions are modified, the minimum mass of accreted ices varies between 13.7 and 18.3 $\mearth$, whereas the corresponding O enrichment ranges between 13.1 and 18.3 times the solar value in the giant planet's atmosphere. In all these cases, the minimum amount of required ices exceeds the mass of heavy elements accreted by the Saturn's formation model of A05b. Moreover, this minimum mass also exceeds the mass of heavy elements present in the planet's atmosphere derived by SG04 (of the order of $\sim 10 \mearth$). However, it is still compatible with the global amount of heavy elements predicted by SG04 ($\leq$ 30~$\mearth$).

\begin{table}[h]
\caption[]{Observed C and N enrichments in Saturn, and calculated enrichments in volatiles in the 
nominal model (CO$_2$:CO:CH$_4$ = 30:10:1 and N$_2$:NH$_3$ = 1). The observed values are taken from Kerola et al. (1997) (a), Flasar et al. (2005) (b) and Briggs and Sackett (1989) (c). Calculated enrichments are calibrated on the observed C enrichment owning the same column label.}
\begin{center}
  \begin{tabular}{lcccc}
            \hline
            \hline
          \noalign{\smallskip}
    Species &  \multicolumn{2}{c}{Observed} & \multicolumn{2}{c}{Calculated}  \\
    & (1) & (2) & (1) & (2)  \\
   \noalign{\smallskip}
  \hline
    \noalign{\smallskip}
    C   &  $3.4 \pm 0.9$$^{~a}$ & $8.8 \pm 1.7$$^{~b}$ &  2.8  & 7.1 \\
    N   &  \multicolumn{2}{c}{$3.1 \pm 0.6$$^{~c}$}         &  2.5  & 6.5 \\
    S   &   &                                                                               & 2.1    &  5.5 \\
    Ar  &    &   								    & 2.0   & 5.1 \\
   Kr  &    &  								    &  2.2  &  5.6 \\
   Xe &   &    								    &  2.6  &  6.7 \\
       \hline
     \end{tabular}
     \end{center}
\label{enrichissements}
\end{table}

\begin{table*}
\caption[]{Minimum mass of accreted water ($M_{\rm water}$), minimum mass of accreted ices ($M_{\rm ices}$) and corresponding minimum O/H abundance (compared to solar value) in Saturn's atmosphere required to fit the observed C enrichments. Calculations are presented for the minimum fits of both KAO C and ground-based N measurements (Kerola et al. 1997; Briggs \& Sackett 1989) and {\it Cassini} C measurement (Flasar et al. 2005) in Saturn's atmosphere. Ranges of CO$_2$:CO:CH$_4$ and N$_2$:NH$_3$ gas-phase ratios used here are those determined by A05c (see text).}

            \begin{center}
         \begin{tabular}[]{llccccc}
            \hline
            \hline
            \noalign{\smallskip}
            &  & \multicolumn{4}{c}{N$_2$:NH$_3$ = 1} & N$_2$:NH$_3$ = 10  \\
            \noalign{\smallskip}
            \cline{3-5}
            \cline{6-7}
             \noalign{\smallskip}
             & & CO$_2$:CO = 4 & CO$_2$:CO = 3 & CO$_2$:CO = 2 & CO$_2$:CO = 1  & CO$_2$:CO = 1   \\
             \noalign{\smallskip}              
             \hline                         
             \noalign{\smallskip}
             KAO & $M_{\rm water} / \mearth$ & 2.2  & 2.5 & 3.0 & 4.0 & 4.6 \\
              & $M_{\rm ices} / \mearth$ &  5.4  & 5.6  & 5.9 & 6.6 & 7.6 \\
              & O/H & 5.1 & 5.4 & 5.8 & 6.6 & 7.6 \\  
             \hline
             \noalign{\smallskip}
            {\it Cassini} & $M_{\rm water} / \mearth$ & 5.7  & 6.6 & 8.0 & 11.0 & 11.2 \\
             & $M_{\rm ices} / \mearth$ &  13.7  & 14.4  & 15.7 & 18.2 & 18.3 \\
             & O/H & 13.1 & 13.9 & 15.2 & 18.0 & 18.3 \\
            \hline
         \end{tabular}
         \end{center}
         \label{ices}
\end{table*}

\section{Discussion and conclusions}

In the framework of the clathrate hydrates trapping theory and using the gas-phase abundances of major volatile species in the solar nebula used by A05c to fit the volatiles enrichments in Jupiter, we have calculated the enrichments in O, C, N, S, Xe, Ar, and Kr with respect to their solar abundances in the atmosphere of Saturn. Two different measurements of the abundance of carbon in Saturn have been considered for the calibration of our calculations, namely the ground-based determination derived by Kerola et al. (1997) and the recent {\it Cassini} measurement as published by Flasar et al. (2005).

Using the near-IR determination of C by Kerola et al. (1997), we are able to fit the measured value of N enrichment. Moreover, the predicted amount of ices accreted by Saturn is consistent with the formation model of A05b and the internal structure model of SG04, provided the minimum I/R ratio required in the giant planet varies between $\sim$ 0.7 and 1.4, a value compatible with the recent {\it Cassini} determination of Phoebe's low density (1630 Kg.m$^{-3}$; Porco et al. 2005).

On the other hand, using the recent far-IR determination of C reported by Flasar et al. (2005), the measured value of N enrichment could no longer be reproduced. In addition, the amount of ices accreted by Saturn exceeds the mass of heavy elements predicted by the A05b model. This difference is likely to be due to the assumptions made by A05b in the calculation of the planetesimal's disk (see also Alibert et al. 2005a for details).

Furthermore, the amount of heavy elements required to explain the {\it Cassini} measurement that should be currently in Saturn's envelope appears incompatible with the results of the internal structure modeling by SG04 (Table \ref{ices}). In addition, in order to explain only the value of C enrichment given by Flasar et al. (2005), one needs to invoke the accretion of at least 10.8 $\mearth$ of ices (including a minimum of 3.9 $\mearth$ of water) in the envelope of Saturn with our nominal gas-phase conditions, an amount slightly exceeding the maximum mass of heavy elements predicted in this zone by SG04. The discrepancy is higher if one adds the amount of ices required to trap N$_2$ and NH$_3$ as clathrate hydrates or hydrates in planetesimals: the measured abondance of N by Briggs \& Sackett (1989) requires as a minimum the additional accretion of 0.9 $\mearth$ of ices, whereas assuming an enrichment of N of 6.5 (as predicted by our calculations) requires the accretion of a minimum of 2.3 $\mearth$ of ices. Therefore, in order to reconcile the Saturn's internal structure models and the C measurement made by the {\it Cassini} spacecraft, one must then suppose a sedimentation of some of the accreted water onto the core, although the major part of other accreted volatiles would have remained in the atmosphere. This assumption is compatible with the work of Fortney and Hubbard (2003; 2004) who argued that a significant oxygen depletion may have occurred in Saturn's atmosphere during its thermal history. This may be due to the immiscibility of oxygen in hydrogen at high temperature and pressure conditions (Fortney and Hubbard 2003).\\

Finally, a number of ground-based C measurements such as those of Buriez \& de Bergh (1981)  and Trafton (1985) are in agreement with the $Cassini$ CIRS data, along with the $Voyager$ IRIS data utilized by Courtin et al. (1984) and the ISO-SWS determination by Lellouch et al. (2001) (see Table \ref{CH4}). This implies that the $Cassini$ determination would appear more representative of the carbon abundance in Saturn's atmosphere than that of Kerola et al. (1997). On this basis, we note that our global oxygen abundance in Saturn is at least $\sim$ twice higher than that predicted by Visscher \& Fegley (2005) from net thermochemical reaction calculations in its atmosphere. On the other hand, the calculations of Visscher \& Fegley (2005) are consistent with the actual Saturn's atmospheric composition and may not be valid for the whole planet.

\begin{acknowledgements}
This work was supported in part by the Swiss National Science Foundation. We thank Jean-Marc Petit and Emmanuel Lellouch for helpful remarks.
\end{acknowledgements}

{}

\clearpage

\end{document}